# Combined selective plane illumination microscopy (SPIM) and full-field optical coherence tomography (FF-OCT) for *in vivo* imaging

RUI MA,[1,2,*] OLGA LYRAKI,[1,2,3] DANIEL WEHNER,[1,2] AND JOCHEN GUCK[1,2,*]

[1]*Max Planck Institute for the Science of Light, Staudtstrasse 2, 91058 Erlangen, Germany*
[2]*Max-Planck-Zentrum für Physik und Medizin, Kussmaulallee 2, 91054 Erlangen, Germany*
[3]*Department of Biology, Friedrich-Alexander-University Erlangen-Nürnberg, 91058 Erlangen, Germany*
*\*rui.ma@mpl.mpg.de*
*\*jochen.guck@mpl.mpg.de*

**Abstract:** Selective plane illumination microscopy (SPIM), also known as light sheet fluorescence microscopy, provides high specificity through fluorescence labeling. However, it lacks complementary structural information from the surrounding context, which is essential for the comprehensive analysis of biological samples. Here, we present a high-resolution, multimodal imaging system that integrates SPIM with full-field optical coherence tomography (FF-OCT), without requiring modifications to the existing SPIM setup. Both SPIM and FF-OCT offer low phototoxicity and intrinsic optical sectioning, making them well-suited for *in vivo* imaging. Their shared detection path enables seamless and efficient co-registration of fluorescence and structural data. We demonstrate the functionality of this combined system by performing *in vivo* imaging of zebrafish larvae.

## 1. Introduction

Selective plane illumination microscopy (SPIM), commonly referred to as light sheet fluorescence microscopy, has become a powerful tool for *in vivo* high-resolution imaging of relatively opaque specimens [1]. Due to its low phototoxicity and 3D optical sectioning capability, it has been widely applied in the fields of developmental biology, neuroscience, and cancer research, allowing researchers to investigate dynamic processes, such as embryogenesis, neuronal activity, and local tumor environments [2–6]. SPIM utilizes a thin sheet of laser beam to excite fluorophores on the focal plane of the detection lens, which is perpendicular to the illumination path, to obtain a two-dimensional image. By eliminating out-of-focus fluorescence excitation using a light sheet, intrinsic non-invasive optical sectioning is achieved with low phototoxicity. A 3D image can be obtained by moving the specimen towards or away from the detection camera and simultaneously acquiring a stack of 2D images. Although SPIM imaging offers high specificity of fluorescence labeling with exceptional contrast, it lacks broader structural context, thus impeding the interpretation of fluorescent structures within the larger anatomical landscape. Providing a detailed anatomical context complements fluorescence data, thereby enhancing the understanding of spatial relationships between biological structures.

Multiple techniques have been combined with SPIM to include structural information while retaining its low phototoxicity and high resolution. Optical tomography [7] utilizes a back-projection algorithm to reconstruct tomographic images by acquiring a stack of image data through spiral specimen movement. This approach overcomes the short depth of field imposed by the relatively high numerical aperture (NA) of the detection objective in SPIM. Three-dimensional bright-field microscopy [8] captures bright-field images from multiple angles, with each angle consisting of a stack of images scanned along the optical axis, which are subsequently fused into a single reconstruction. Both optical tomography and three-dimensional bright-field microscopy thereby complement light sheet microscopy to provide high-resolution structural context. However, both techniques require sample rotation and rely heavily on computational reconstruction, making them time-consuming due to the need for precise sample mounting and the additional time required for computational analysis. This complexity can hinder real-time observations and make high-throughput studies challenging. Additionally, the use of trans-illumination results in low-contrast images for translucent biological specimens. Alternatively, a swept-source optical coherence tomography (SS-OCT) [9] using reflected illumination has been combined with digital scanned light sheet fluorescence microscopy (LSFM). However, tunable lasers for SS-OCT operate at wavelengths longer than 1 µm, limiting its spatial resolution. For example, the lateral resolution of SS-OCT reported by Khajavi et al. 2021 is approximately 14.9 µm, thereby restricting the system's ability to resolve fine or subcellular structures in biological samples. Furthermore, such systems require expensive components, including tunable lasers and high-speed detectors, sophisticated optical configurations, and complex image processing algorithms, making SS-OCT impractical to integrate with conventional SPIM, where beam steering is not required.

Here, we introduce time-domain based full-field optical coherence microscopy (FF-OCT) [10,11], which complements SPIM without requiring any modification to the SPIM setup. FF-OCT offers intrinsic optical sectioning capability, low phototoxicity, and does not require complex image acquisition and reconstruction procedures, all of which factors make this technique well-suited for *in vivo* imaging applications. Additionally, the LED utilized in FF-OCT has a central wavelength of 565 nm and a relatively broad bandwidth, providing a high lateral resolution of approximately 0.75 µm and a high axial resolution of less than 2 µm. Moreover, SPIM and FF-OCT share a common detection path, allowing for seamless co-registration of fluorescence signals with structural information. We demonstrate the functionality of the combined SPIM-FF-OCT setup by applying it to live zebrafish larvae.

## 2. Methods and materials

Fig. 1 illustrates a simplified schematic of the combined SPIM-FF-OCT setup, and Visualization 1 provides further details on the combined setup. The illumination paths of the SPIM and FF-OCT are indicated in green and red, respectively. The two imaging modalities share the same detection path, which consists of a tube lens and a camera. A flippable quadband emission filter is inserted in the detection path during SPIM image acquisition.

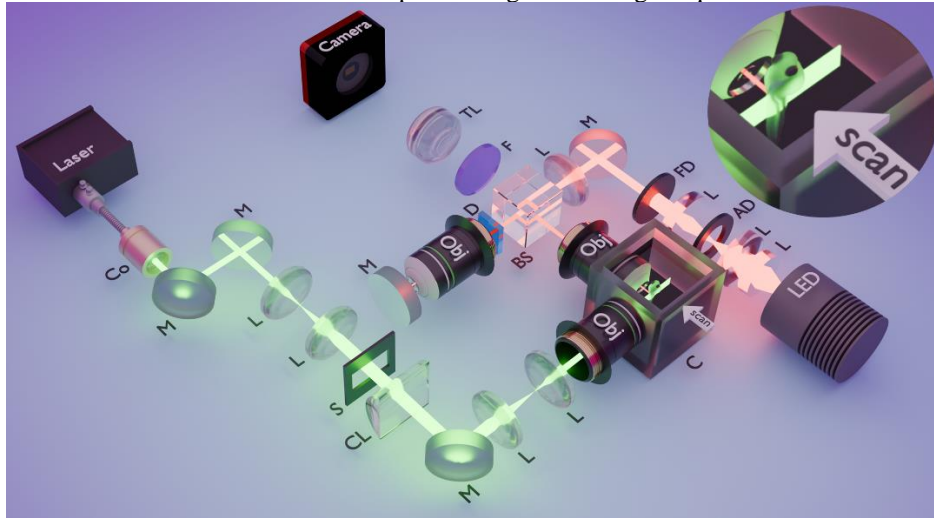

Fig. 1. Schematic of the combined SPIM-FF-OCT setup. The green beam represents the SPIM illumination path, and the red beam indicates the illumination path of the FF-OCT. Note that the objects shown are not to scale. Inset:

Magnified view of the imaging chamber part including mounted sample. The sample is mounted on a 4D stage and scanned towards or away from the camera. Abbreviations: Co, fiber collimator; M, mirror; L, lens; S, vertical slit; CL, cylindrical lens; C, sample chamber filled with water; FD, field diaphragm; AD, aperture diaphragm; BS, beam splitter cube; D, dispersion compensator; F, Quadband emission filter (flippable); TL, tube lens; Obj, objective lens.

## 2.1 SPIM

We used a laser combiner (C-Flex C8, HÜBNER Photonics GmbH, Germany) with multiple laser lines as the excitation light source. The laser was collimated with a triplet apochromatic fiber collimator (60FC-4-RGBV11-47, Schäfter + Kirchhoff GmbH, Germany). The selection of the laser lines was digitally controlled by an Arduino Uno microcontroller board. Optical components were chosen as reported by the OpenSPIM platform to create a diffraction-limited light sheet that is projected onto the focal plane of the detection objective, with a 10× objective lens (UMPLFLN10XW, Olympus, Germany) for illumination and a 20× objective lens for detection (UMPLFLN20XW, Olympus, Germany) [12,13]. A Quadband Emitter (F72-866, Semrock) was used as an emission filter. For multi-channel fluorescence imaging, the combination of a laser combiner and a quadband filter not only speeds up the process but also simplifies the SPIM setup without the need for a motorized filter wheel or additional cameras. A tube lens (U-TLU-1-2, Olympus) and a CMOS camera (CS165MU1/M, Thorlabs, Germany) were used for the shared detection path.

The sample was mounted on motorized linear stages, where the XY stages (L-509.10AD10, PI) and a Z stage (L-509.10DG10, PI) were controlled by a C-884.4DC controller. Both illumination and detection objective lenses and the sample were immersed in aqueous fish medium held by an imaging chamber. 3D images were acquired by scanning the sample towards the camera with the Z stage via Multi-Dimensional Acquisition (MDA), a built-in user interface in µManager [14].

## 2.2 FF-OCT

As white light source, we used an epi-Köhler illumination scheme for FF-OCT with a 565-nm LED of 104 nm bandwidth (M565L3, Thorlabs, Germany). A 455-nm longpass filter (FGL455M, Thorlabs, Germany) was added immediately after the LED to filter out the side peak around 420 nm in order to obtain a quasi-Gaussian spectrum. A non-polarizing beamsplitter cube (BS013, Thorlabs, Germany) split the beam into the reference arm and the sample arm, which was shared with SPIM. The reference arm had the same 20× objective lens as the sample arm. In addition, it had an OD 6.0 filter acting as the reflective mirror in order to match the intensity of the weak, back-scattering signal from the sample. A dispersion compensator was inserted in the reference arm to compensate for the water-induced dispersion from the sample arm.

To retrieve the amplitude and phase information at a given depth, commonly N-phase shifting algorithms (N ≥ 3) are used, where N consecutive images are phase shifted by $2\pi/N$ by moving the reference arm with a step size of $\frac{\lambda_0}{2n \cdot N}$ ($\lambda_0$: center wavelength of the light source, n: refractive index of the immersion medium). However, adding a motorized stage on the reference arm increases the cost and complicates the software design. Although our z stage has a limited minimum step size, it is able to perform 2-phase shifting method. Despite the fact that 2-phase shifting method is not able to decouple amplitude and phase, and the retrieved 2D image at a certain depth $\frac{I(x,y,\varphi_0) - I(x,y,\varphi_0 + \pi))}{2}$ is a contribution of amplitude and phase: $A(x,y) cos\varphi_0$, we found that this method is sufficient for retrieving 2D structural information. Additionally, due to the low coherence length LED, 3D optical sectioning can be obtained and out-of-focus signals excluded.

Due to the low bit-depth and moderate full well capacity of the CMOS camera, each phase shift was averaged 1000 times to achieve a reasonable SNR.

## 2.3 Zebrafish husbandry and transgenic lines

Zebrafish were maintained and reared in a research fish facility (Tecniplast, Italy) under a 14/10 h light/dark cycle according to FELASA recommendations [15,16]. We used larval zebrafish aged up to 120 hours post-fertilization (hpf), derived from voluntarily mating adult zebrafish (permit: I/39/FNC of the Amt für Veterinärwesen und gesundheitlichen Verbraucherschutz Stadt Erlangen). Procedures on zebrafish larvae up to 120 hpf is not regulated as animal experiments by the European Commission Directive 2010/63/EU. We used *elavl3*:GFP-F$^{mps10}$ transgenic zebrafish larvae in which all neurons are labeled with membrane-tethered GFP [17]. Embryos were treated with 0.00375 % 1-phenyl-2-thiourea (P7629, Sigma-Aldrich) beginning at 24 hpf to prevent pigmentation.

### 2.4 Zebrafish mounting for imaging

Zebrafish larvae were anesthetized in E3 medium [15] (henceforth referred to as fish medium) containing 0.02% Ethyl 3-aminobenzoate methanesulfonate (MS-222; Tricaine PharmaQ, PharmaQ) and transferred to 1% Ultra-Pure$^{TM}$ Low Melting Point agarose (16520, Invitrogen) in fish medium. Zebrafish larvae were drawn into a FEP tube (BOHLS1815-04, Bohlender) with an inner diameter of 0.8 mm, using an 18G blunt needle attached to a 1 ml syringe. The syringe body was mounted on the sample holder attached to the motorized stages. The sample chamber was filled with 0.01% MS-222-containing fish medium to keep preparations from drying out.

## 3. Results

### 3.1 Spatial resolution of the SPIM-FF-OCT system

Green-emission fluorescent microspheres with a diameter of 0.5 µm, embedded in 1% low-melting-point agarose, were used to determine the lateral resolution of the combined system. Fig. 2(a) shows the fluorescent image of the beads with a full width at half maximum (FWHM) of 0.75 µm. Since the emission wavelength of the fluorescent microsphere is around 520 nm, very close to the center wavelength of the 565-nm LED used for the FF-OCT component, the measured lateral resolution applied to both SPIM and FF-OCT.

To measure the axial resolution of the SPIM, we used a glass plate rotated 45° towards the illumination and detection path so that the light sheet with a wavelength of 488 nm was reflected perpendicular to the camera. We scanned the plate parallel to the detection path with a range of 800 µm at 10 µm intervals and reconstructed the beam waist, finding that the thinnest part of the light sheet is 4.07 µm (Fig. 2b).

We next performed an autocorrelation function measurement (Fig. 2c) to determine the axial resolution of the FF-OCT setup. A slightly tilted glass plate was mounted on the sample arm to create interference lines. We then scanned the glass plate for a total distance of 5.5 µm with 0.1 µm intervals. The fitted FWHM of the autocorrelation function is 2.0 µm in air. For biological imaging, the samples were immersed in fish medium, and the corresponding axial resolution is 1.5 µm.

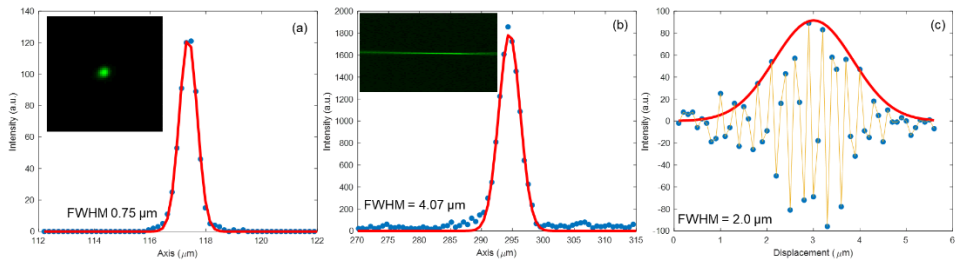

Fig. 2. (a) Lateral resolution measurement of the combined SPIM-FF-OCT system. Inset shows an image of a fluorescent microsphere. (b) Axial resolution of the SPIM. Inset shows: image of the SPIM light sheet beam waist. (c)

Autocorrelation measurement using a 565 nm LED, corresponding to an axial resolution of 2.0 µm in air for the FF-OCT setup.

### 3.2 Co-registered fluorescence and structural imaging of live zebrafish larvae

To demonstrate the functionality of the combined SPIM and FF-OCT system, we performed *in vivo* imaging of zebrafish larvae, which are ideal for SPIM application due to their optical transparency. We used *elavl3*:GFP-F transgenic zebrafish in which all neurons are labelled with membrane-tethered GFP. Fig. 3 shows FF-OCT and SPIM images of the trunk of a two-day-old (days post-fertilization; dpf) zebrafish. Using the combined setup, we were able to distinguish between different anatomical structures, including the spinal cord, notochord, and muscle by FF-OCT (Fig. 3a). While the current setup is not able to fully image highly pigmented regions of the zebrafish trunk (indicated by the white arrows), since the signal in these areas reaches the maximum readout limit of the camera, this limitation can be improved by employing cameras with higher bit-depth and full well capacity. The GFP fluorescence image of the spinal cord was obtained from the same plane by SPIM (Fig. 3(b)). By obtaining both SPIM and FF-OCT data, the resulting images containing fluorescence and structural information, respectively, could be combined to link the two information types (Fig. 3c). Fig. 3(d) shows the field of view (FOV) in the context of the zebrafish body. The FOV of our combined setup is 244.8 µm × 183.6 µm, which is determined by the number of pixels of the utilized camera (1440 ×1080) and can be further extended by integrating cameras with more pixels to cover the entire width of the zebrafish trunk.

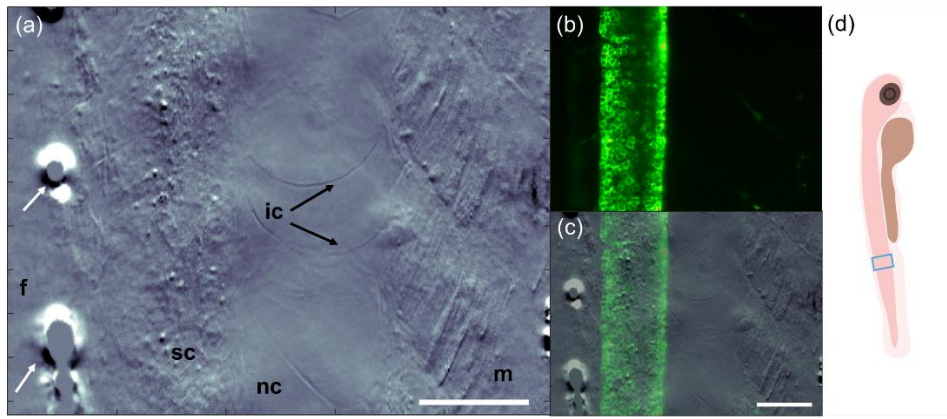

Fig. 3. FF-OCT and SPIM images of the trunk of a 2 dpf *elavl3*:GFP-F transgenic zebrafish (rostral is up, dorsal s left). The fluorescently labelled neurons of the spinal cord (GFP) are indicated in green. (a) FF-OCT image (white arrows indicate highly pigmented regions). (b) SPIM image. (c) Co-registration of SPIM and FF-OCT images. (d) Field of view in the context of the zebrafish larva. Abbreviations: f, median finfold; sc, spinal cord; nc, notochord; m, muscle; ic, interior cell of the notochord (black arrows). Scale bar: 50 µm.

### 3.3 Optical sectioning with combined FF-OCT and SPIM

We next explored the optical sectioning capability of the combined system by imaging a comparable trunk region of zebrafish at 2 dpf and 4 dpf (Fig. 4 and Fig. 5). In the 2 dpf zebrafish, clear structural changes were observed over a scanning interval of 10 µm (Fig. 4). Beyond providing fluorescence data on neuronal distribution within the spinal cord region, the combined system also revealed surrounding structures that exhibited distinct variations with depth. Such depth-dependent structural differences were also evident in the 4 dpf larva (Fig. 5), where the FOV was slightly adjusted towards the median finfold while retaining the fluorescent spinal cord region to better demonstrate the 3D sectioning capability of FF-OCT. Due to the thinness of the finfold, depth scanning progressively brought it into focus, while neurons distributed across a width of over 40 µm to ~ 10 µm showed a gradual shift in their centers towards the notochord region. In addition, structural changes of the skeletal muscle could be observed between the developmental stages analyzed.

Zebrafish larvae have become a prime model for spinal cord injury research due to their elevated regenerative capacity for the central nervous system [18]. Previous studies often employed bright-field microscopy to visualize tissue backgrounds as a complement to fluorescent information [17,19,20]. As shown in Fig. 4 and Fig. 5, FF-OCT provides high-resolution structural background information with micron-level optical sectioning, offering a detailed reference framework to complement fluorescence signals.

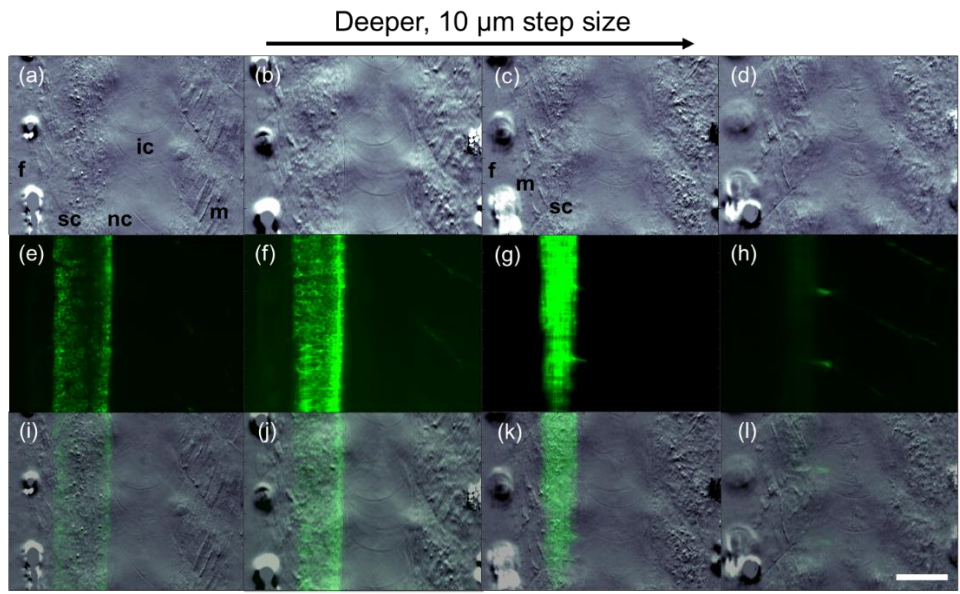

Fig. 4. FF-OCT and corresponding SPIM images of the trunk of a 2 dpf *elavl3*:GFP-F transgenic zebrafish (rostral is up, dorsal s left) with 10 µm scan interval, demonstrating the optical sectioning capability of the combined setup. The fluorescently labelled neurons of the spinal cord (GFP) are indicated in green. (a)-(d): FF-OCT images. (e)-(h): corresponding SPIM images. (i)-(l): Co-registered FF-OCT and SPIM images. Abbreviations: f, median finfold; sc, spinal cord; nc, notochord; ic, interior cells of the notochord; m, muscle. Scale bar: 50 µm.

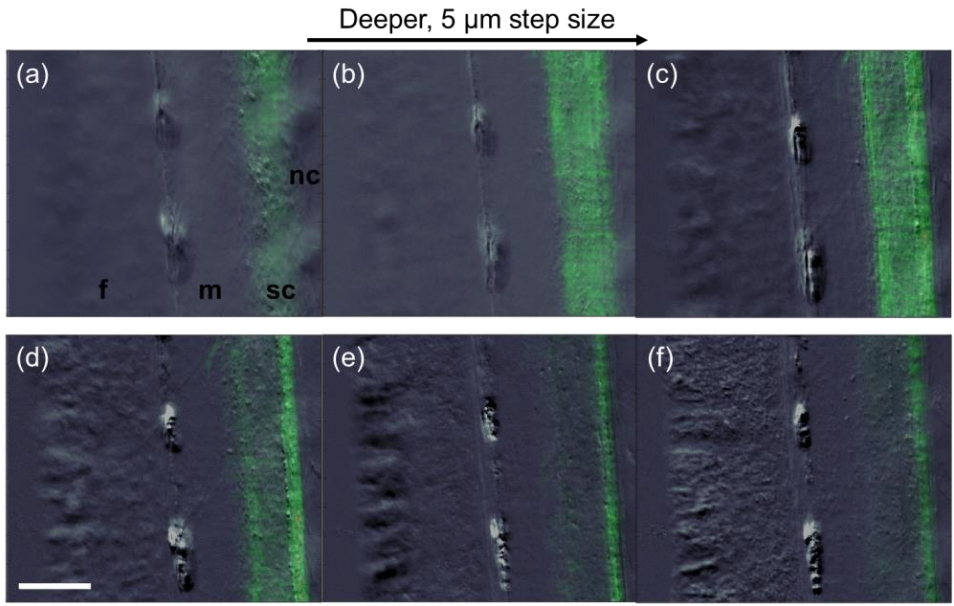

Fig. 5. Co-registered FF-OCT and SPIM images of the trunk of a 4 dpf *elavl3*:GFP-F transgenic zebrafish (rostral is up, dorsal s left) with 5 µm scan interval. The fluorescently labelled neurons of the spinal cord (GFP) are indicated in green. Abbreviations: f, median finfold; m, muscle; sc, spinal cord; nc, notochord. Scale bar: 50 µm.

## 4. Conclusions

We have demonstrated a multimodal, non-invasive imaging system that integrates SPIM with FF-OCT for *in vivo* imaging. By incorporating FF-OCT into a SPIM setup without the need for additional modifications, we propose a practical and viable addition to existing SPIM systems. This combined setup is able to provide both fluorescent and structural information with seamless co-registration through a shared detection path. Moreover, the image acquisition and processing pipeline is straight forward, eliminating the need for beam steering or complex image reconstruction algorithms. We used this system to visualize spatial and temporal variations in internal structures of the trunk of live zebrafish at two different developmental stages, demonstrating the potential of this multimodal approach for *in vivo* imaging applications.

This combined system could be further developed to perform 3D whole-body time-lapse imaging, which is currently impractical due to the need for extensive averaging. This is primarily due to the low bit depth and full well capacity of the camera, but could be overcome by replacing the current camera with advanced CMOS cameras equipped with passive or liquid cooling, which could reduce image acquisition time by 2- to 3-fold. This improvement, combined with image stitching technique [21] would thus allow for detailed studies of whole-organism development. Moreover, analyzing temporal changes of dynamic structures in static FF-OCT images, namely dynamic FF-OCT (D-FF-OCT) [22–24], enables the visualization of weakly back-scattered structures at the subcellular level based on intracellular motility and metabolic activity. This capability offers new insights into differentiating cell types, monitoring cell proliferation, and tracking migration *in vivo*, and its integration with SPIM would provide still more detailed visualization of biological systems.

The combined SPIM and FF-OCT system we report here is well-suited for visualizing live biological samples with minimal phototoxicity. Integrating structural and fluorescence data enables precise and comprehensive analysis of biological samples by providing both detailed spatial context and functional insights. This approach allows for a deeper understanding of complex biological phenomena while minimizing ambiguities that might arise from using either modality independently. The combined system also enables the visualization of the spatial and temporal relationships between anatomic structures and biological functions, leading to more informed conclusions and enhancing the interpretation of complex biological processes.

**Acknowledgments.** We gratefully acknowledge Dr. Catherine Xu for her critical review of the manuscript. We thank Casandra Cecilia Carrillo Mendez and Olga Stelmakh for excellent fish care. The authors acknowledge financial support from the Max Planck Society (to J. Guck) and from the Deutsche Forschungsgemeinschaft (project number 492010287 – SPP 2332 Physics of Parasitism to J. Guck; project number 460333672 - CRC1540 Exploring Brain Mechanics to J. Guck and D. Wehner).

**Disclosures.** The authors declare no conflicts of interest.

**Data availability.** Data underlying the results presented in this paper are not publicly available at this time but may be obtained from the authors upon reasonable request.